\documentstyle[preprint,aps]{revtex}
\begin{document}
\draft
\preprint{SNUTP 96-054}
\vskip 0.3cm
\title{Quantum Description of Anyons:
Role of Contact Terms}
\author{Sang-Jun Kim \\and \\Choonkyu Lee}
\address{Department of Physics and Center for
Theoretical Physics, Seoul National University,
Seoul,151-742,Korea}
\maketitle
\begin{abstract}
We make an all-order analysis to establish the
precise correspondence between nonrelativistic
Chern-Simons quantum field theory and an 
appropriate first-quantized description.
Physical role of the field-theoretic contact
term in the context of renormalized perturbation
theory is clarifed through their connection to
self-adjoint extension of the Hamiltonian in the
first-quantized approach. Our analysis provides
a firm theoretical foundation on quantum field
theories of nonrelativistic anyons.
\end{abstract}
\newpage
\renewcommand{\theequation}
{\arabic{section}.\arabic{equation}}
\section{INTRODUCTION}
It is generally understood that for every Schr\"{o}dinger
(i.e., nonrelativistic) quantum field theory, there exists
a corresponding quantum mechanical(or first-quantized) 
description\cite{fw}. This will certainly be the 
case with non-singular interactions, but becomes quite
uncertain if singular interactions, such as the two-body 
$\delta$-function potential, are involved. In two or more
spatial dimensions, the $ \delta $-function potentials
lead to nontrivial (i.e., interacting) systems only when
infinite renormalization or self-adjoint extension of the
Hamiltonian is taken into consideration \cite{bf,ag,jb}.
Using the language of Schr\"{o}dinger field theory, on
the other hand, the $ \delta $-function potential between
particles is formally represented by a local or contact 
interaction of the form $ (\phi^{\dagger}(x)\phi(x))^2. $
The Dyson-Feynman perturbation theory is then beset with
ultraviolet divergences ---to make sense out of this field
theory, one must regularize and renormalize the amplitudes.
Because of these complications it is not so easy to make
a direct comparison between the first- and second-
quantized approaches even for this relatively simple 
system.
But, Bergman \cite{ob} demonstrated recently that the two
approaches are in fact completely equivalent once the
renormalized strength of the $(\phi^{\dagger}(x)\phi(x))^2$
interaction is chosen to be related in a specific way to
the self-adjoint extension parameter entering the quantum
mechanical approach.

In two spatial dimensions, we have another kind of system
where similar problems arise naturally---those involving
anyons \cite{lm}. Quantum mechanically, anyons can be
regarded as flux-charge composites and so the relative
dynamics of the two anyon system is essentially the
Aharonov-Bohm scattering problem \cite{ab}. The latter
problem has certain ambiguity as regards the choice of
boundary condition at the singular point of the 
Aharonov-Bohm potential, namely, at the location of the
flux line. According to the theory of self-adjoint 
extension, it is known that there exist a one-parameter
family of acceptable boundary conditions \cite{ag,mt},
including as a special case the often-assumed hard-core
boundary condition(appropriate to an impenetrable
flux line)\cite{ab}. In perturbative treatments, however,
this boundary condition is not easily incorporated and
without paying due attention to it one ends up with a
divergent perturbation series\cite{elf}. Among various
proposals made to amend this situation\cite{cc}, a 
particularly interesting one is to introduce an extra
$\delta$-function potential of suitable strength\cite{ga},
the role of which is to secure a finite(and correct)
perturbation theory. The second-quantized description
has certain ambiguity also. Naively, the field theory for
anyons can be obtained if the Schr\"{o}dinger field is
coupled to a Chern-Simons gauge field \cite{crh,jp}.
But, once one takes renormalization into account, one is
forced to allow the $(\phi^{\dagger}(x)\phi(x))^2$-type
local interaction also in the theory\cite{bl}. Needless to
say, it should be important to understand the physical role
of such contact interaction term and also its significance
within the first-quantized approach. This is especially
so in view of possible relevance of the Chern-Simons
field theory in some of most remarkable phenomena in
planar physics such as the quantum Hall effect.

A serious study on the above issue was made recently by
Amelino-Camelia and Bak \cite{ac}. They showed that 
lower-order field theory calculations of the scattering
amplitudes agree with the corresponding quantum 
mechanical results obtained using the method of 
self-adjoint extension, provided the strength of the 
$(\phi^{\dagger}(x)\phi(x))^2$-interaction is suitably
related to the self-adjoint extension parameter. Note that
this situation is entirely analogous to the pure 
$\delta$-function case mentioned earlier. A new feature
is that, if the strength of the 
$(\phi^{\dagger}(x)\phi(x))^2$-interaction is equal to
the critical value appropriate to the so-called self-dual
limit\cite{jp}, the field theory turns out to be 
ultraviolet finite\cite{bl} and yields the anyon 
scattering amplitude consistent with the scale-invariant
boundary conditions. This last fact has now been confirmed
by one of us to {\it all} orders in perturbation theory
\cite{sk}. With these developments it should be useful to
have a self-contained account on quantum description of
anyons which takes contact terms properly into 
consideration, in both first- and second- quantized 
formulations. This will, above all, serve to show the
essential equivalence of the two formulations and 
illuminate the physical significance of the contact term
entering either formulation. The present work has been
written precisely for this purpose.

This paper is organized as follows. In Sec. \ref{se:2}
we reconsider quantum mechanical description of anyons and
explain in particular how the boundary condition, needed
at the coincidence point of two anyons, can be implemented
through the introduction of appropriate contact terms in
the Hamiltonian. Note that we here consider not just the
hard-core boundary condition but the general boundary
condition allowed by the self-adjoint extension method
[The corresponding anyons were called `colliding anyons'
in Ref.\cite{ac}]. 
 Sec. \ref{se:3} is concerned with the Chern-Simons 
field theory description of anyons, with the
$(\phi^{\dagger}(x)\phi(x))^2$-type contact terms 
included for the sake of renormalizability. In this field
theory context we elaborate on the all-order analysis of
the two-anyon s-wave scattering amplitude, a brief account
of which was given for the first time in Ref. \cite{sk}.
Based on  this analysis, we then clarify the connection
between the field theoretic formulation and the quantum
mechanical description. Sec. \ref{se:4} contains the
summary and discussions.

\section{FIRST-QUANTIZED DESCRIPTIONS OF ANYONS WITH 
BOUNDARY-CONDITION IMPLEMENTING CONTACT TERMS}
\label{se:2}
\setcounter{equation}{0}
Anyons, which are realized only in planar physics, can
be described using either bosonic or fermionic
description. Taking bosonic description, one may specify
quantum dynamics of a system of anyons by the particle
Hamiltonian
\begin{eqnarray}
H&=&\sum_n \frac{1}{2m}({\bf p}_{n} -\alpha {\bf A}_{n})^2
+\frac{1}{2}\sum_{n,m(\neq n)} U(|{\bf r}_{n}-
{\bf r}_{m}|), \label{eq:21}\\
A_{n}^{i}&\equiv&\epsilon^{ij}\sum_{m(\neq n)}\frac{
x_{n}^{j}-x_{m}^{j}}{|{\bf r}_{n}-{\bf r}_{m}|^2},
\label{eq:22}
\end{eqnarray}
where ${\bf r}_n \equiv (x_{n}^{1},x_{n}^{2})$ denotes
the position of the $n$-th particle, and $ U(|{\bf r}_{n}
-{\bf r}_{m}|)$ a non-singular two-body potential included
for generality. The vector potential ${\bf A}_{n}\equiv
(A_{n}^{1},A_{n}^{2})$ seen by particle $n$ is that of point
vortices carried by all the other particles, and it is the
resulting Aharonov-Bohm-type interference effect that is
responsible for the anyonic behavior of particles.
The parameter $ \alpha $, called `statistical parameter',
characterizes the type of anyons (i.e., their statistics) and
without loss of generality $\alpha$ may be restricted to
the interval ($-1$,1).

But, due to the singular nature of the vector potential
(\ref{eq:22}) at points ${\bf r}_{n}={\bf r}_{m}$, the
information given above does {\it not} specify the system
completely --- a suitable boundary condition at locations
of singularity must be posited. There exist a class of
boundary conditions (see below) all of which are in fact
realizable with the help of suitable regularization
procedures. Thus, anyons are further classified by the
nature of the boundary condition chosen at two-anyon
intersection points. Also this extra specification is
something that {\it has to be made} and should be 
taken into account, say, in all approximate treatments.
Divergences encountered in naive perturbation theory
(with $|\alpha| \ll 1$) for the anyon system can be
ascribed to the ill-defined nature of the problem caused
by not fixing the boundary condition\cite{cc}.
Under the circumstance that the boundary condition cannot
be ignored, it will then be natural to ask whether the
boundary condition in question may be implemented by having
instead an appropriate contact interaction term in the
Hamiltonian. That is possible, as we will show below.
This entirely-Hamiltonian description for anyons will
allow a straightforward application of perturbation theory
and also serve as a bridge to the field-theoretic 
description.

To study the boundary condition problem mentioned above,
we may concentrate on the relative dynamics in the 
two anyon sector. It will be governed by the Hamiltonian
\begin{eqnarray}
H_{rel}&=&\frac{1}{m}({\bf p}-\alpha {\bf A})^2+U(r),
\label{eq:23}\\
A^{i}&\equiv&\frac{\epsilon^{ij}x^{j}}{r^2},
\hskip 20pt
(r\equiv |{\bf r}|) \label{eq:24}
\end{eqnarray}
where ${\bf r}\equiv(x^{1},x^{2})$ and {\bf p} denote
relative position and momentum. Furthermore, since the
boundary condition to be chosen at $ r=0$ has an effect
on the s-wave only, it will suffice to consider the
s-wave Hamiltonian. Throughout this paper, we also set
$ \hbar = 1. $ The corresponding Schr\"{o}dinger
equation reads 
\begin{equation}
\left[-\frac{1}{r}\frac{d}{dr}r\frac{d}{dr}+\frac{
\alpha^2}{r^2}+mU(r)\right]\psi(r)=k^2\psi(r),
\hskip 20pt (r>0)
\label{eq:25}
\end{equation}
which has the general solution of the form (for $ \alpha
\neq 0$)
\begin{equation}
\psi(r)=A {\cal J}_{|\alpha|}(kr)+B
{\cal J}_{-|\alpha|}(kr),
\label{eq:26} 
\end{equation}
where A, B are arbitrary constants.
Here, ${\cal J}_{\pm |\alpha|}(kr)$ represent two linearly 
independent solutions of Eq. (\ref{eq:25}) with the
following small-$r$ behaviors
\begin{equation}
{\cal J}_{\pm|\alpha|}(kr)\sim
\frac{1}{\Gamma(1\pm|\alpha|)}\left(\frac{kr}{2}\right)
^{\pm |\alpha|}[1+{\cal O}(r)],
\hspace{15pt}r\rightarrow 0.
\label{eq:27}
\end{equation}
Note that we have assumed a sufficient regularity of $ U(r)$
at $ r=0$, and the functions ${\cal J}_{\pm |\alpha|}(kr)$
have been normalized in such a way that they reduce to 
ordinary Bessel functions, i.e., $ J_{\pm |\alpha|}(kr)$
if $U(r)$ is taken to vanish. 
The wave function in Eq. (\ref{eq:26}) is
not regular at $ r=0$, but still square integrable (with
$|\alpha|<1$) for arbitrary finite values of A and B.
This is in marked contrast with the case of higher partial
waves where only one type of solution is square integrable.
In the present case we may pick an appropriate boundary
condition at the origin in accordance with the method of
self-adjoint extension of the Hamiltonian.
This leads to a one-parameter family of boundary
condition \cite{ag,mt}
\begin{equation}
\lim_{r\rightarrow 0}\left[r^{|\alpha|}\psi_{\theta}({
\bf r})-\frac{\tan\theta}{\mu^{2|\alpha|}}
\frac{\Gamma(1+|\alpha|)}{\Gamma(1-|\alpha|)}\frac{
d}{d\left(r^{2|\alpha|}\right)}\left(r^{|\alpha|}\psi_{\theta}
({\bf r})\right)\right]=0
\label{eq:28}
\end{equation}
with the corresponding solution given as
\begin{equation}
\psi_{\theta}(r)=(const.)\left[{{\cal J}_{|\alpha|}(kr)+
\tan\theta\left(\frac{k}{\mu}\right)^{2|\alpha|}{\cal J}
_{-|\alpha|}(kr)}\right],
\label{eq:29}
\end{equation}
where $\theta$ is a dimensionless real parameter and $\mu$
a reference scale. The hard-core boundary condition (i.e.,
$ \psi (0)=0$), which is often assumed in the literature,
corresponds to the choice $ \theta=0$ but, clearly, there is
no a priori reason to favor this particular case.

We now discuss how we can implement the boundary condition
(\ref{eq:28}) {\em dynamically}, {\em viz}., by introducing
a contact interaction term of the form
\begin{equation}
H_c=\lambda_{\theta}(r)\delta^{2}({\bf r})
\label{eq:210}
\end{equation}
in the Hamiltonian. [In the notation of Eq. (\ref{eq:21}),
this contact term translates into two-body interaction of
the form $ H_c=\frac{1}{2}\sum_{n,m(\neq n)}\lambda_{
\theta}(|{\bf r}_n -{\bf r}_m |)\delta^2 ({\bf r}_n
-{\bf r}_m ). $]
In Eq. (\ref{eq:210}) we have written a $\delta$-function
multiplied by an $r$-dependent function, an apparently
redundant expression. But, as we shall see soon, it has a
reason.\footnote{This is related to the fact that strength
of the $\delta$-function in our case needs
renormalization\samepage
(except for the special cases of $ \theta=0$ or $ \frac{\pi}
{2}$). The situation here is analogous to the case of a 
pure $\delta$-function potential problem, as discussed in 
Refs. \cite{bf,ag,jb}.} If we form the new relative
Hamiltonian
\begin{equation}
\tilde{H}_{rel}=H_{rel}+H_{c},
\label{eq:211}
\end{equation}
the function $ \psi_{\theta}$ given in Eq. (\ref{eq:29})
will then have to be its eigenstate(when all ill-behaved 
quantities are interpreted in a suitably regularized sense).
This demands especially that
\begin{equation}
\langle\psi|\tilde{H}_{rel}|\psi_{\theta}\rangle=
\frac{k^2}{m}\langle\psi|\psi_{\theta}\rangle
\label{eq:212}
\end{equation}
or
\begin{equation}
\int d^2 {\bf r} \psi^{\ast}(r)\left\{\frac{1}{m}(-i
{\bf \nabla}-\alpha{\bf A})^{2}+U(r)+\lambda_{\theta}
(r)\delta^{2}({\bf r})-\frac{k^2}{m}\right\}
\psi_{\theta}(r)=0,
\label{eq:213}
\end{equation}
where $ \psi(r)$ can be an arbitrary function of the form
(\ref{eq:26}). One must choose $ \lambda_{\theta}(r)$ such
that Eq. (\ref{eq:213}) may hold. In using the condition
(\ref{eq:213}) it is to be noted that, since both $ \psi^{
\dagger}(r)$ and $ \psi_{\theta}(r) $ are in general not
regular at $ r=0 $, the operation with the
$ \delta $-function term demands much care and especially
one may not insert $ \lambda_{\theta}(0) $ for
$ \lambda_{\theta}(r). $ Actually, there are other
ill-behaved contributions in Eq. (\ref{eq:213}) also.
See below.

To give a precise meaning to Eq. (\ref{eq:213}) and also
to the Schr\"{o}dinger equation with 
$ \tilde{H}_{rel} $ (in the range $ r\geq 0$), we must 
introduce a suitable
regularization procedure as regards the singularity at $ r
=0. $ This may be effected by replacing the vector
potential (\ref{eq:24}) by the regularized
expression
\begin{equation}
A^{i(\epsilon)}=\frac{1}{r+\epsilon}\left(\frac{\epsilon^{
ij}x^{j}}{r}\right)
\label{eq:214}
\end{equation}
and the $ \delta $-function $ \delta^{2}({\bf r})$ by the
regularized one
\begin{equation}
\delta^{2}({\bf r};\epsilon)=\frac{1}{2\pi}{\bf \nabla}^2
\log(r+\epsilon)=\frac{1}{2\pi r}\frac{\epsilon}{(r+
\epsilon)^2 },
\label{eq:215}
\end{equation}
with $ \epsilon \rightarrow 0+$ understood if there is no
further dangerous manipulation left. 
Our regularization procedure is simply
to replace $ r $ by $ r+\epsilon $, while leaving all
angular dependences intact. Then it is possible
to show that the functions obtained from ${\cal J}_{\pm|
\alpha|}(kr)$ by the simple substitution $ r\rightarrow
r+\epsilon $ satisfy the equations (for $ r\geq 0$)
\begin{eqnarray}
&&\left[{\frac{1}{m}\left(-i{\bf \nabla}-\alpha{\bf 
A}^{(\epsilon)}\right)^{2}+U(r)-\frac{k^2}{m}}\right]
{\cal J}_{\pm |\alpha|}(k(r+\epsilon))
\nonumber \\
&&=\left[\frac{1}{m}\right. \left(-\frac{1}{r}\frac{d}{dr}
r\frac{d}{dr}+\frac{\alpha^2 }{(r+\epsilon)^2 }\right)+U(r)
-\left.\frac{k^2}{m}\right]{\cal J}_{\pm |\alpha|}(k(r+
\epsilon))\nonumber \\
&&=\mp\frac{|\alpha|}{m}\frac{\left[\frac{k(r+\epsilon)}
{2}\right]^{\pm |\alpha|}}{\Gamma(1\pm|\alpha|)}\frac{
\epsilon}{r(r+\epsilon)^2} \nonumber \\
&&\equiv\mp\frac{2\pi|\alpha|}{m}\frac{\left[\frac{k
(r+\epsilon)}{2}\right]^{\pm |\alpha|}}{\Gamma(1\pm
|\alpha|)}\delta^{2}({\bf r}; \epsilon).
\label{eq:216}
\end{eqnarray}
These follow readily from the behaviors (\ref{eq:27}) and
the fact that ${\cal J}_{\pm|\alpha|}(kr)$ solve the 
differential equation (\ref{eq:25}). Using Eq.
(\ref{eq:216}) we may thus conclude that the function
$ \psi_{\theta}(r+\epsilon)$, given by the expression
(\ref{eq:29}) with the substitution $ r\rightarrow
r+\epsilon $, satisfies the relation
\begin{eqnarray}
&&\left[{\frac{1}{m}\left(-i{\bf \nabla}-\alpha{\bf 
A}^{(\epsilon)}\right)^{2}+U(r)-\frac{k^2}{m}}\right]
\psi_{\theta}(r+\epsilon) \nonumber\\
&&=-\frac{2\pi|\alpha|}{m}\left\{{\frac{
\frac{1}{\Gamma(1+|\alpha|)}\left[\frac{\mu(r+\epsilon)}{2}
\right]^{|\alpha|}-\tan\theta\frac{1}{\Gamma(1-|\alpha|)}
\left[\frac{\mu(r+\epsilon)}{2}\right]^{-|\alpha|}
}{\frac{1}{\Gamma(1+|\alpha|)}\left[\frac{\mu(r+\epsilon)}{2}
\right]^{|\alpha|}+\tan\theta\frac{1}{\Gamma(1-|\alpha|)}
\left[\frac{\mu(r+\epsilon)}{2}\right]^{-|\alpha|}}
}\right\}\delta^{2}({\bf r}; \epsilon)\psi_{\theta}(r+
\epsilon).
\label{eq:217}
\end{eqnarray}
Note that what we have on the right hand side of this 
equation can yield a nontrivial contribution, say, to
the matrix element formed with the general state given in
Eq. (\ref{eq:26}).

From Eq. (\ref{eq:217}) the precise form of the necessary
contact interaction term can be inferred: the quantity
multiplying $ \psi(r+\epsilon)$ in the right hand side of
Eq. (\ref{eq:217}) should be identified with 
$ -\lambda_{\theta}(r)\delta^{2}({\bf r}) $ or, in
a regularized form, with $-\lambda_{\theta}(r)\delta^{2}
({\bf r};\epsilon). $ That is\footnote{If one wishes,
the form given in Eq. (\ref{eq:218}) for $ \lambda_{
\theta}$ may be replaced by another expression involving
only the regularization parameter $ \epsilon $ but not
$ r. $ [One may use the integral condition (\ref{eq:213})
for this purpose.] But, with a regularized
$ \delta $-function brought in, there is no reason to
favor such expression in particular; indeed,
our development leads quite naturally to the form
(\ref{eq:218}).},
\begin{eqnarray}
\lambda_{\theta}(r)&=&-\frac{2\pi|\alpha|}{m}\left\{{
\frac{\Gamma(1+|\alpha|)\tan\theta-\Gamma(1-|\alpha|)
\left[\frac{\mu(r+\epsilon)}{2}\right]^{2|\alpha|}
}{\Gamma(1+|\alpha|)\tan\theta+\Gamma(1-|\alpha|)
\left[\frac{\mu(r+\epsilon)}{2}\right]^{2|\alpha|}}
}\right\} \nonumber \\
&=&\left[
\begin{array}{l}
\displaystyle
\frac{2\pi|\alpha|}{m},\hskip 20pt \theta=0
\vspace{12pt}\\
\displaystyle
-\frac{2\pi|\alpha|}{m}\left\{{1-2\frac{\Gamma(1-|\alpha|)}
{\Gamma(1+|\alpha|)}\cot\theta\left[\frac{\mu(r+\epsilon)}
{2}\right]^{2|\alpha|}}\right\}, \theta \neq 0.
\end{array}
\right.
\label{eq:218}
\end{eqnarray}
With the thus-constructed contact term included in $ \tilde
{H}_{rel} $, this new Hamiltonian, without any separate
consideration of the boundary condition at the origin, will
select the function-type $ \psi(r)=\psi_{\theta}(r)$(i.e.,
the $ \epsilon\rightarrow 0+ $ limit of
$ \psi_{\theta}(r+\epsilon) $) as its only acceptable 
eigenfunction type. Especially, the hard-core boundary 
condition (i.e., the $ \theta=0 $ case) is
implemented by a particularly simple contact Hamiltonian, $
H_c=\frac{2\pi|\alpha|}{m}\delta^2 ({\bf r}). $
This repulsive contact term is precisely what has been
suggested by the authors of Ref. \cite{ga} as the extra
interaction needed to cancel perturbation-theory divergences.
Also interesting is the fact that, since
\begin{eqnarray}
B &\equiv&\epsilon_{ij}\partial_{i}A_{j}^{(\epsilon)}
\nonumber \\
&=&-\frac{\epsilon}{r(r+\epsilon)^2 }=-2\pi\delta^2 ({\bf
r};\epsilon),
\label{eq:219}
\end{eqnarray}
the net regularized relative Hamiltonian
$ \tilde{H}_{rel} $ (see Eq. (\ref{eq:211}) in the
$ \theta=0 $ case assumes the form of the (spin-fixed)
two-dimensional Pauli Hamiltonian withthe vector potential
$ {\bf A}^{(\epsilon)} $ and the scalar potential
$ U(r). $

Note that, even for the same given boundary condition, the
contact Hamiltonian may have a slightly different look if
the method of regularization is different. For example,
suppose we replace the vector potential (\ref{eq:24}) by
that of finite-radius flux tube of radius $ r_0 $ (with the
magnetic field confined to the surface of the tube)
\cite{crprl}, i.e. by
\begin{eqnarray}
A^{i(r_0 )}=\left[\begin{array}{c}
0, \hskip 25pt r<r_0
\vspace{6pt} \\
\displaystyle\frac{\epsilon^{ij}x^{j}}{r^2},
\hskip 10pt r>r_0
\end{array}
\right. \label{eq:220}
\end{eqnarray}
Then, for the regularized form of the contact
Hamiltonian, the ring potential
\begin{equation}
H_c =\frac{\bar{\lambda}_{\theta}}{2\pi r_{0}}\delta(r-r_0 
) \hskip 25pt \left(\approx \bar{\lambda}_{\theta}\delta^2 
(r)\right) \label{eq:221}
\end{equation}
will especially be appropriate. 
Here the strength $ \bar{\lambda}_{
\theta}$ is to be chosen such that the boundary condition
(\ref{eq:28}) may be realized for $ r_0 \rightarrow 0. $
Using this form, Hagen\cite{cr} demonstrated how one may
implement the hard-core boundary condition. This can easily
be generalized to accomodate more general boundary condition
in Eq. (\ref{eq:28}). Now the eigenfunction of $ H_{rel} $
will have the form (up to an overall multiplicative
constant)
\begin{eqnarray}
\psi(r)=\left[\begin{array}{l}
{\cal J}_{0}(kr), \hskip 25pt r<r_0 \\
A{\cal J}_{|\alpha|}(kr)+B{\cal J}_{-|\alpha|}(kr), r>r_0
\end{array}
\right. \label{eq:222}
\end{eqnarray}
with the constants $A$ and $B$ determined by the conditions
\begin{eqnarray}
\psi(r_0 +\epsilon)-\psi(r_0-\epsilon)&=&0, \nonumber \\
\left.\frac{d\psi}{dr}\right|_{r=r_0 +\epsilon}-\left.
\frac{d\psi}{dr}\right|_{r=r_0 -\epsilon}&=&m\frac{\bar{
\lambda}_{\theta}}{2\pi r_0}\psi(r_0 ).
\label{eq:223}
\end{eqnarray}
For sufficiently small $kr_0$, Eq. (\ref{eq:223}) imply
\begin{eqnarray}
A\frac{1}{\Gamma(1+|\alpha|)}\left(\frac{kr_0 }{2}\right)^
{|\alpha|}+B\frac{1}{\Gamma(1-|\alpha|)}\left(\frac{kr_0 }
{2}\right)^{-|\alpha|}=1, \nonumber \\
A\frac{|\alpha|}{\Gamma(1+|\alpha|)}\left(\frac{kr_0 }{2}
\right)^{|\alpha|}-B\frac{|\alpha|}{\Gamma(1-|\alpha|)}
\left(\frac{kr_0 }{2}\right)^{-|\alpha|}=\frac{m}{2\pi}
\bar{\lambda}_{\theta}, \label{eq:224}
\end{eqnarray}
and hence we obtain the ratio
\begin{equation}
\frac{B}{A}=\frac{\Gamma(1-|\alpha|)\left(|\alpha|-\frac{
m\bar{\lambda}_{\theta}}{2\pi}\right)}{\Gamma(1+|\alpha|)
\left(|\alpha|+\frac{m\bar{\lambda}_{\theta}}{2\pi}\right)}
\left(\frac{kr_0 }{2}\right)^{2|\alpha|}.
\label{eq:225}
\end{equation}
On the other hand, if we compare the above wave function in
the region $r>r_0$ with the form given in Eq. (\ref{eq:29}),
we are led to set
\begin{equation}
\frac{B}{A}=\tan\theta\left(\frac{k}{\mu}\right)^{
2|\alpha|}. \label{eq:226}
\end{equation}
From Eq. (\ref{eq:225}) and Eq. (\ref{eq:226}) we thus see
that the strength $ \bar{\lambda}_{\theta} $ of the given
contact term should be chosen as
\begin{equation}
\bar{\lambda}_{\theta}=-\frac{2\pi|\alpha|}{m}
\frac{\Gamma(1+|\alpha|)\tan\theta-\Gamma(1-|\alpha|)
\left(\frac{\mu r_0 }{2}\right)^{2|\alpha|}}
{\Gamma(1+|\alpha|)\tan\theta+\Gamma(1-|\alpha|)
\left(\frac{\mu r_0 }{2}\right)^{2|\alpha|}}.
\label{eq:227}
\end{equation}
Note that, for $ \theta=0 $, we again find the value
$ \bar{\lambda}_{\theta=0}=\frac{2\pi|\alpha|}{m}. $

We have so far shown that an anyon system can be specified
solely by the Hamiltonian, only when one takes into account
a suitable contact interaction term. It is needed to
implement the boundary condition chosen at the two-particle
intersection point. For the special case of $ \theta=0 $ or
$ \frac{\pi}{2} $, this contact Hamiltonian assumes a
particularly simple form, {\em viz}., $ H_c =\frac{2\pi|
\alpha|}{m}\delta^2 ({\bf r})$ for $ \theta=0 $  and 
$ H_c =-\frac{2\pi|\alpha|}{m}\delta^2({\bf r}) $ for
$ \theta=\frac{\pi}{2}. $ In fact, only for $ \theta=0 $ or
$ \theta=\frac{\pi}{2}$, $H_c$ and also $\psi_{\theta}(r)$
in Eq. (\ref{eq:29}) (up to an irrelevant overall constant)
become independent of our reference scale $\mu$; this is
an evidence of the {\em scale invariance} in the system.
Here one might suspect that, for a general $ N $-anyon 
Hamiltonian, contact interaction terms involving more than
two particles may have to be introduced as well. We strongly
believe that these should be unnecessary, i.e., two-body
contact interactions we have discussed suffice. This is
supported by the perturbative analysis of an N-anyon system
(in Ref.\cite{ga}, for example) and also by the
renormalization counterterm structure in the field theoretic
approach.

Before closing this section, we will give the explicit
expression for the s-wave scattering amplitude of two
anyons when the two-body potential $ U({\bf r}_n -{\bf r}
_m )$ is taken to be zero. Following the analysis given
in Ref.\cite{ab} yields the amplitude
\begin{equation}
A_s (p)=(e^{-i\pi|\alpha|}-1)\frac{(\mu/p)^{2|\alpha|}-\tan
\theta}{(\mu/p)^{2|\alpha|}+e^{-i\pi|\alpha|}\tan\theta},
\label{eq:228}
\end{equation}
where $ p $ is the magnitude of the relative
momentum.
Defining
\begin{equation}
\lambda_{ren} \equiv -\frac{4\pi|\alpha|}{m}
\frac{\tan\theta-1}{\tan\theta+1}, \label{eq:229}
\end{equation}
this may be rewritten as
\begin{eqnarray}
A_s(p)
&=& e^{-i\pi|\alpha|}-1 -(e^{i\pi|\alpha|}-e^{-i\pi
|\alpha|})\frac{\lambda_{ren}-\lambda_0}{\lambda_0-
\lambda_{ren}+(\lambda_0+\lambda_{ren})
(\mu/p)^{2|\alpha|}e^{i\pi|\alpha|}} \nonumber \\
&=& e^{-i\pi|\alpha|}-1 -\frac{e^{i\pi|\alpha|}-e^{
-i\pi|\alpha|}}{2\lambda_0}(\lambda_{ren}-\lambda_0)
\nonumber \\&&+\frac{e^{i\pi|\alpha|}-e^{
-i\pi|\alpha|}}{2\lambda_0}\frac{\frac{(\mu/p)^{
2|\alpha|}e^{i\pi|\alpha|}-1}{2\lambda_0}
}{
1+(\lambda_{ren}+\lambda_0)\frac{(\mu/p)^{2|\alpha|
}e^{i\pi|\alpha|}-1}{2\lambda_0}}(\lambda_{ren}^2-
\lambda_0^2),
\label{eq:230}
\end{eqnarray}
where $ \lambda_0 = \frac{2\pi|\alpha|}{m}. $
[Note that the appropriate expression with the hard-core
boundary condition, as adopted in Ref. \cite{ab},
is obtained if we set $ \lambda_{ren}=\lambda_0 $ in
Eq. (\ref{eq:230}).] We may
now of course look on this result as that corresponding
to the system defined by the Hamiltonian (\ref{eq:211})
(with $ U\equiv 0 $) which contains the contact interaction.
We will make use of this result in the next section.

\section{QUANTUM FIELD THEORETIC DESCRIPTION OF ANYONS}
\label{se:3}
\setcounter{equation}{0}
When interactions involved are nonsingular, a
Schr\"{o}dinger quantum field theory is known to be
completely equivalent to nonrelativistic quantum mechanics
of many particles\cite{fw}. Needless to say, Feynman 
diagram approach in many body theory is an important
byproduct of this correspondence. But, in the presence of
local or contact interactions, the singular nature of
interaction makes the situation no longer simple---both 
infinite renormalization (in the field theoretic approach)
and self-adjoint extension of the Hamiltonian (in the
quantum mechanical approach) take parts in any discussion 
purporting to establish the analogous correspondence.
In the latter case, we are not aware of any general
argument as regards the nature of such correspondence
and so each model system has to discussed separately. [The
main obstacle to giving {\em general} arguments stems from
the big difference in the flavored language for the two
approaches, one diagrammatical (and in momentum space) and
the other in the form of differential equations (in 
position space)]. The anyon system involves singular
interactions and this issue arises naturally. We will 
below give a field theory description of anyons
(including renormalization effects) and then relate it
to the quantum mechanical description of the
previous section. For earlier related works, see Refs.
\cite{ob}, \cite{bl} and \cite{ac}.

We begin by specifying our candidate quantum field theory
for anyons. It is a (2+1)-D nonrelativistic system 
described by the Lagrangian density
\begin{equation}
{\cal L}= \frac{\kappa}{2}{\partial}_t {\bf A} \times
{\bf A}- \kappa A_0 B+\phi^\dagger (i D_t +
\frac{{\bf D}^2}{2m})
\phi - \frac{\lambda}{2} \phi ^\dagger \phi ^\dagger
\phi \phi ,
\label{eq:31}
\end{equation}
where $ \phi $ is a bosonic field, {\bf A}=($ A_1 $,$ A_2 $)
denotes a Chern-Simons gauge field, $ B $=$ \epsilon_{ij}
\partial_iA_j \equiv{\bf \nabla}\times {\bf A}$ and the 
covariant derivatives are
\begin{eqnarray}
D_t &=& \partial _t +i e A_0   \nonumber \\
{\bf D} &=& {\bf \nabla}-i e {\bf A}.
\label{eq:32}  
\end{eqnarray}
Without the last term in Eq. (\ref{eq:31}), this model was 
first considered by Hagen\cite{crh}, But the last contact
interaction term, first considered in Ref. \cite{jp}, is
necessary to ensure the renormalizability of the theory.
As it turns out, this additional term is of crucial
importance in the field theoretic treatment of anyons.
For a comprehensive review on various aspects concerning
the above theory, readers may consult Ref. \cite{gd}.

Setting aside the renormalization problem for the moment,
it might be useful to reproduce quantum mechanical 
description corresponding to the above theory by the
standard many-body-theory procedure. First, using the
`Gauss' constraint
\begin{equation}
{\bf \nabla} \times {\bf A}=-\frac{e}{\kappa}\phi^{\dagger}
\phi, \label{eq:33}
\end{equation}
the gauge fields $ {\bf A} $ may be expressed in terms
of the matter fields (in the Coulomb gauge) as
\begin{equation}
{\bf A}({\bf r},t)=-\frac{e}{\kappa}{\bf \nabla}\times\int
d^2 r' G({\bf r}-{\bf r}')\phi^{\dagger}({\bf r}',t)\phi
({\bf r}',t),
\label{eq:34}
\end{equation}
where $ G({\bf r}) $ is the Green's function of the
two-dimensional Laplacian
\begin{equation}
G({\bf r})=\frac{1}{2\pi}\ln |{\bf r}|. \label{eq:35}
\end{equation}
Assuming Eq. (\ref{eq:34}), the Hamiltonian can now be 
identified with
\begin{equation}
H=\int d^2 r\left[\frac{1}{2m}({\bf D}\phi)^{\dagger}
\cdot({\bf D}\phi)+\frac{\lambda}{2}\phi^\dagger
\phi^\dagger \phi \phi\right], \label{eq:36}
\end{equation}
Then, defining the $N$-particle Schr\"{o}dinger wave
function
\begin{equation}
\Phi({\bf r_1},\cdots,{\bf r_N},t)\equiv\frac{1}{N!}
\left\langle{0{\left|\phi({\bf r}_1 ,t)\cdots
\phi({\bf r}_N ,t)\right|}\Phi}\right\rangle 
\label{eq:37}
\end{equation}
and using the canonical commutation relations
\begin{equation}
[\phi({\bf r},t),\phi({\bf r}',t')]=
[\phi^{\dagger}({\bf r},t),\phi^{\dagger}({\bf r}',t')
]=0,
\hskip 10pt
[\phi({\bf r},t),\phi^{\dagger}({\bf r}',t')]=
\delta({\bf r}-{\bf r}'), \label{eq:38}
\end{equation}
it is straightforward (but tedious) to derive the 
Schr\"{o}dinger equation of the form
\cite{jp,cy}
\begin{eqnarray}
i\frac{\partial}{\partial t}\Phi({\bf r_1},\cdots,
{\bf r_N},t)&=&\left\{ {-\frac{1}{2m}\sum_n \left[ {
{\bf \nabla}_n -\frac{ie^2}{\kappa}{\bf \nabla}_n
\times\left({\sum_{m(\neq n)}G({\bf r}_n-{\bf r}_m)}
\right)}\right]^2}\right. \nonumber\\ &+& \left.{
\frac{\lambda}{2}\sum_{n,m(\neq n)}
\delta({\bf r}_n -{\bf r}_m )}\right\}
\Phi({\bf r_1},\cdots,{\bf r_N},t). \label{eq:39}
\end{eqnarray}

What we have in Eq. (\ref{eq:39}) has the appearance of the
Schr\"{o}dinger equation for the anyon system, with
$ \alpha=\frac{e^2}{2\pi\kappa}$ and $ U(|{\bf r}_n-
{\bf r}_m|)=0 $ in the notation of Sec. \ref{se:2};
the $\delta$-function potential in Eq. (\ref{eq:39}), which
originates from the $ \phi^\dagger \phi^\dagger 
\phi \phi $ coupling in Eq. (\ref{eq:31}), may be viewed as
the boundary-condition implementing term at two-particle
intersection points. But the above argument suggests at
most the formal correspondence for {\em bare} amplitudes
(with singular interactions replaced by suitably 
regularized one, as we have done in Sec. \ref{se:2}).
Our goal is to find the correspondence between well-defined
renormalized amplitudes of the two approaches.
In the quantum mechanical description, we have invoked
the method of self-adjoint extension to find such
well-defined two-particle scattering amplitude which
depends on the self-adjoint extension parameter $ \theta $
(or on $ \lambda_{ren}$, as defined in Eq. (\ref{eq:229}))
but not on the bare contact coupling $ \lambda. $
To be able to make an unambiguous comparison, the 
corresponding renormalized amplitude in the field theory
context will be obtained below. Here it is perhaps 
worthwhile to remark that the correspondence we found 
above (for bare
quantities) may be more than a formal one if the theory is
free from ultraviolet divergences; this happens for the
$ \lambda $-value equal to $ \pm\frac{2\pi|\alpha|}{m}$,
for which we have a scale-invariant system. In this
connection, see the paragraph following Eq. (\ref{eq:227})
and the discussion following immediately after 
Eq. (\ref{eq:339}) below.

Given the Lagrangian density (\ref{eq:31}), Feynman rules
are as follows. The nonrelativistic boson propagator in
momentum space is
\begin{eqnarray}
\Delta(k)=
\frac{1}{\left(k_0 -\frac{{\bf k}^2}{2m}
+ i\epsilon\right)}, \label{eq:310}
\end{eqnarray} 
so that we have\footnote{Here and also in
Eq. (\ref{eq:313}), one usually has free-field operators
$ \phi^f $ and $ A^f $ instead of full fields $ \phi $
and $ A $. But, because of the reason to be explained
shortly, there is no mass and field renormalization in
our theory so that the free-field restriction can be
omitted.}
\begin{equation}
-i\langle 0|T \phi(x) \phi^{*}(0)|0\rangle =
\int \frac{d^3 k}{(2\pi)^3} \Delta(k)
e^{-i(k_0 x_0-{\bf k}\cdot{\bf x})}. \label{eq:311}
\end{equation}
Introducing the gauge fixing term
\begin{equation}
{\cal L}_{gf}=-\frac{1}{\xi}({\bf \nabla}\cdot{\bf A})^2
\label{eq:312}
\end{equation}
and then considering the limit $ \xi \rightarrow 0 $, the
only nonvanishing components of the Chern-simons gauge
field-propagator is easily found to be
\begin{equation}
-i\langle 0|T A_i (x) A_0 (0)|0\rangle =
\int \frac{d^3 k}{(2\pi)^3} D_{i0}(k)e^{-i(k_0 x_0-
{\bf k}\cdot{\bf x})}. \label{eq:313}
\end{equation}
with
\begin{equation}
D_{i0}(k)=-D_{0i}(k)=\frac{i
\epsilon^{ij}k_j }{\kappa {\bf k}^2}, \label{eq:314}
\end{equation}
There are four interaction vertices as shown in Fig.1.
Three of them coming from the covariant derivative terms,
are given as
\newcounter{var}
\addtocounter{equation}{1}
\setcounter{var}{\value{equation}}
\renewcommand{\theequation}
{\arabic{section}.\arabic{var}\alph{equation}}
\setcounter{equation}{0}
\begin{eqnarray}
\Gamma_0&=&-ie,\\
\Gamma_i&=&\frac{ie}{2m}(p_i +p'_i),\\
\Gamma_{ij}&=&-\frac{ie^2}{m}\delta_{ij},
\end{eqnarray}
\setcounter{equation}{\value{var}}
\renewcommand{\theequation}
{\arabic{section}.\arabic{equation}}
while the other from the contact interaction term reads
\begin{equation}
\Gamma_{\lambda}=-2i\lambda.
\end{equation}

With our gauge choice (that is, the Coulomb gauge) and 
normal ordering of contact interaction term $ \frac
{\lambda}{2} \phi^\dagger \phi^\dagger \phi \phi $,
diagrams in Fig.2 give zero contribution. So there are
no renormalization of mass, field and charge $ e $ in 
this nonrelativistic theory. It is only the strength
of contact interaction which has a nontrivial
renormalization effect. This requires a detailed study of
two-particle scattering amplitude.

The two-particle scattering amplitude can be described in
terms of the effective two-particle interaction\cite{fw}
$ \Gamma $, represented by (see Fig.3)
\begin{eqnarray}
\Gamma(p,p';q,q')&=&K(p,p';q,q')+\int\frac{d^3 k}
{(2\pi)^3}K(p,p';q,q')i\Delta(k)i\Delta(p+p'-k)
\nonumber \\
&&\cdot K(k,p+p'-k;q,q')+\cdots, \label{eq:317}
\end{eqnarray}
where $K$ denotes the two-particle-irreducible kernel.
The {\em entire} nonvanishing graphs which are not
reducible by cutting two matter lines are those shown
in Fig.4, and the full kernel $ K $ can be identified with
the sum of $ K_a $, $ K_b $ and $ K_c . $
[Note that, in this nonrelativistic field theory, graphs
like those shown in Fig.5 vanish indentically.]

Using the Feynman rules, one then finds that the quantities
$ K_a $, $ K_b $ and $ K_c $ are given by the following
expressions (which are independent of energy variables):
\addtocounter{equation}{1}
\setcounter{var}{\value{equation}}
\renewcommand{\theequation}
{\arabic{section}.\arabic{var}\alph{equation}}
\setcounter{equation}{0}
\begin{eqnarray}
K_a (p,p';q,q')&=& \frac{e^2}{m\kappa} \frac{({\bf q}-{
\bf p})\times({\bf p}-{\bf p}')}{({\bf p}-{\bf q})^2}
\label{eq:318a}
\\
K_b (p,p';q,q')&=& -i\frac{e^4}{4\pi m\kappa^2}\ln\frac{
\Lambda^2}{({\bf p}-{\bf q})^2} \label{eq:318b}\\
K_c (p,p';q,q')&=& -i\lambda
\label{eq:318c}
\end{eqnarray}
\setcounter{equation}{\value{var}}
\renewcommand{\theequation}
{\arabic{section}.\arabic{equation}}
\hspace {-13pt}
In Eq. (\ref{eq:318b}), $ \Lambda $ is an ultraviolet
momentum cutoff and one can obtain the given 
expression as
\begin{eqnarray}
K_b (p,p';q,q')&=&-i\frac{e^4}{2\pi m\kappa^2}\int
\frac{d^3 k}{(2\pi)^3}\frac{i}{p_0 +k_0 -\frac{({\bf p}
+{\bf k})^2}{2m}+i\epsilon} \hskip 4pt
\frac{{\bf k\cdot}({\bf q}-{\bf p}-{\bf k})}{{\bf k}^2 
({\bf q}-{\bf p}-{\bf k})^2} \nonumber \\
&&+ ({\bf p}\rightarrow {\bf p}',{\bf q}
\rightarrow{\bf q}')\nonumber\\
&=&-i\frac{e^4}{4\pi m\kappa^2}\int\frac{
d^2 {\bf k}}{(2\pi)^2}\frac{{\bf k\cdot}({\bf q}-{\bf p}
-{\bf k})}{{\bf k}^2 ({\bf q}-{\bf p}-{\bf k})^2} +
({\bf p}\rightarrow {\bf p}',{\bf q}\rightarrow{\bf q}')
\nonumber \\
&=&-i\frac{e^4}{4\pi m\kappa^2}\int_0 ^{2\pi} d\varphi
\int_0 ^\Lambda \frac{|{\bf k}|d|{\bf k}|}{(2\pi)^2}
\frac{|{\bf k}||{\bf q}-{\bf p}|\cos\varphi -{\bf k}^2}{
{\bf k}^2({\bf k}^2 -2|{\bf k}||{\bf q}-{\bf p}|\cos
\varphi+({\bf q}-{\bf p})^2)} \nonumber \\
&&+({\bf p}\rightarrow {\bf p}',{\bf q}\rightarrow{\bf q}')
\nonumber \\
&=& -i\frac{e^4}{4\pi m\kappa^2}\ln\frac{
\Lambda^2}{({\bf p}-{\bf q})^2} 
\end{eqnarray}
Also note that the equation (\ref{eq:317}) for $ \Gamma $
may be recast as the integral equation
\begin{eqnarray}
\Gamma(p,p';q,q')&=&K(p,p';q,q')+\int\frac{d^3 k}{(2\pi)^3}
K(p,p';k,p+p'-k)\frac{i}{k_0 -\frac{{\bf k}^2}{2m}+
i\epsilon}\nonumber \\
&&\cdot\frac{i}{p_0 +p_0 '-k_0-\frac{({\bf p}+{\bf p}'-
{\bf k})^2}{2m}+i\epsilon}\Gamma(k,p+p'-k;q,q'), 
\label{eq:320}
\end{eqnarray}
which is the Bethe-Salpeter equation. From
Eq. (\ref{eq:320}) we see that $ \Gamma $ depends not on
$ p_0 $ or $ p_0 ' $ separately but on the sum 
$ p_0 +p_0 ' $ only.
This allows one to perform the $ k_0 $-intergration
immediately (using Cauchy's theorem), to yield the equation
\begin{eqnarray}
\Gamma(p,p';q,q')&=&K(p,p';q,q')+\int\frac{d^2 {\bf k}}{
(2\pi)^2}K(p,p';k,p+p'-k) \nonumber \\
&&\cdot\frac{i}{p_0 +p_0 '- \frac{
{\bf k}^2}{2m}-\frac{({\bf p}+{\bf p}'-{\bf k})^2}{2m}+
i\epsilon}\Gamma(k,p+p'-k;q,q'). \label{eq:321}
\end{eqnarray}

We find it convenient to work in the center of mass frame
where
\begin{equation}
{\bf p}'=-{\bf p}, \hskip 15pt
{\bf q}'=-{\bf q}, \hskip 15pt
p_0 +p_0 '=q_0 +q_0 '
\equiv E.
\end{equation}
Then the simplified notations $ \Gamma\rightarrow\Gamma(
{\bf p},{\bf q};E)$, $ K\rightarrow K({\bf p},{\bf q})$,
etc. for the corresponding quantities should suffice, and
Eq. (\ref{eq:321}) becomes
\begin{equation}
\Gamma({\bf p},{\bf q};E)=K({\bf p},{\bf q})+\int\frac{
d^2 {\bf k}}{(2\pi)^2}K({\bf p},{\bf k})\frac{i}{E-\frac{
{\bf k}^2}{m}+i\epsilon}\Gamma({\bf k},{\bf q};E). 
\label{eq:323}
\end{equation}
If we now decompose $ \Gamma $ and $ K $ as
\begin{eqnarray}
\Gamma({\bf p},{\bf q};E)=\sum_n \Gamma^n (|{\bf p}|,
|{\bf q}|,E) e^{in\varphi}, \hskip 10pt
K({\bf p},{\bf q})=\sum_n K^n (|{\bf p}|,
|{\bf q}|) e^{in\varphi}
\end{eqnarray}
($ \varphi $ is the angle between the incoming and outgoing
momenta) and insert these into Eq. (\ref{eq:323}), it
follows after the angle integration that
\begin{equation}
\Gamma^n (|{\bf p}|,|{\bf q}|,E)=K^n (|{\bf p}|,|{\bf q}|)
+\int\frac{d |{\bf k}|^2}{4\pi}K^n (|{\bf p}|,|{\bf k}|)
\frac{i}{E-\frac{{\bf k}^2}{2m}+i\epsilon}\Gamma^n
(|{\bf k}|,|{\bf q}|,E). \label{eq:325}
\end{equation}
The $ n $-th partial wave part of $ \Gamma $ being obtained
by iterating $ K^n $ only, we are entitled to consider each
partial wave contribution separately. As discussed in Ref.
\cite{sk}, non-s-wave (i.e., $ n\neq 0 $) parts can be
shown to be finite order by order while the series for
$ \Gamma^0 $ (obtained by iterating $ K^0 $) is not.
This in turn implies that renormalization is necessary only
for the s-wave amplitude $ \Gamma^0 . $
So, for our purpose (i.e., to compare the field theoretic
results with those of Sec. \ref{se:2}), it should
suffice from now on to confine our attention to the
analysis of the s-wave amplitude, that is, to the
$ n=0 $ case with the intergral equation (\ref{eq:325}).
Note that, from Eqs. (\ref{eq:318a})-(\ref{eq:318c}),
we have the s-wave contribution of the kernel given as
\addtocounter{equation}{1}
\setcounter{var}{\value{equation}}
\renewcommand{\theequation}
{\arabic{section}.\arabic{var}\alph{equation}}
\setcounter{equation}{0}
\begin{eqnarray}
K^0 &=& K^0_a+K^0_b+K^0_c ,\\
K^0_a&=&0,\\
K^0_b&=&-i\frac{e^4}{4\pi m\kappa^2}\ln
\frac{\Lambda^2}{L(|{\bf p}|,|{\bf q}|)^2},
\label{eq:326c}\\
K^0_c&=&-i\lambda \label{eq:326d},
\end{eqnarray}
\setcounter{equation}{\value{var}}
\renewcommand{\theequation}
{\arabic{section}.\arabic{equation}}
\hspace {-13pt}
where $ L(|{\bf p}|,|{\bf q}|) $ denotes the larger of
$ |{\bf p}| $ and $ |{\bf q}|. $

We will organize the series obtained by iterating the
integral equation for $ \Gamma^0 $ in the following
way. Let $ {\bar{\Gamma}}^0 $, $ \Gamma_{cross} $ and
$ \Gamma_{bubble} $ denote the contributions given
schematically by
\begin{eqnarray}
{\bar{\Gamma}}^0&=&K^0_b + \int K^0_b (i\Delta)(i\Delta)
K^0_b + \int K^0_b (i\Delta)(i\Delta)K^0_b(i\Delta)
(i\Delta)K^0_b + \cdots, \label{eq:327}\\
\Gamma_{cross}&=&K^0_c + \int K^0_c (i\Delta)(i\Delta)
K^0_b + \int K^0_b (i\Delta)(i\Delta)K^0_c +
\int K^0_c (i\Delta)(i\Delta)K^0_b (i\Delta)(i\Delta)
K^0_b \nonumber\\
&&+ \int K^0_b (i\Delta)(i\Delta)K^0_c (i\Delta)
(i\Delta)K^0_b + \int K^0_b (i\Delta)(i\Delta)K^0_b 
(i\Delta)(i\Delta)K^0_c + \cdots, \label{eq:328}\\
\Gamma^0_{bubble}&=&\int K^0_c (i\Delta)(i\Delta)K^0_c +
\int K^0_c (i\Delta)(i\Delta)K^0_b (i\Delta)(i\Delta)
K^0_c \nonumber\\
&&+\int K^0_c (i\Delta)(i\Delta)K^0_b (i\Delta)
(i\Delta)K^0_b (i\Delta)(i\Delta)K^0_c + \cdots,
\label{eq:329}
\end{eqnarray}
respectively\footnote{The quantity $ \bar{\Gamma}^0 $
given by Eq. (\ref{eq:327}) was denoted as
$ \Gamma_{AB}^0 $ in Ref. \cite{sk}, for this
amplitude was identified somewhat mistakenly with the
s-wave part of the Aharonov-Bohm (AB) amplitude obtained
under the hard-core boundary condition in that paper.}. 
Denoting $ K^0_b $ by the graph shown in
Fig.6, these three amplitudes can be expressed
graphically as in Fig.7. Then it is not difficult to
see that the full s-wave amplitude $ \Gamma^0 $ is given
by
\begin{eqnarray}
\Gamma^0={\bar{\Gamma}}^0 + \Gamma_{contact}
\label{eq:330}
\end{eqnarray}
with
\begin{eqnarray}
\Gamma_{contact}&=&\lambda\tilde{\Gamma}_{cross}\left\{
{1+\lambda\tilde{\Gamma}_{bubble}+
\left({\lambda\tilde{\Gamma}_{bubble}}\right)^2
+ \cdots}\right\} \nonumber \\
&=&\frac{\lambda\tilde{\Gamma}_{cross}}{1-\lambda\tilde
{\Gamma}_{bubble}}, \label{eq:331}
\end{eqnarray}
where we have defined $ \tilde{\Gamma}_{cross}
\equiv\frac{1}{\lambda}
\Gamma_{cross} $ and $ \tilde{\Gamma}_{bubble}\equiv
\frac{i}{\lambda^2}\Gamma_{bubble}. $
Note that all contact interaction terms are included in
the quantity $ \Gamma_{contact} $ (as defined by
Eq. (\ref{eq:331})). On the other hand,
$ {\bar{\Gamma}}^0 $ has no dependence on the contact
coupling $ \lambda. $

Let us now look into the cutoff dependence of the
quantities $ {\bar{\Gamma}}^0 ,\tilde{\Gamma}_{cross} $
and $ \tilde{\Gamma}_{bubble}. $ Writing $ x=\Lambda^2,
\alpha=\frac{e^2}{2\pi\kappa} $ and using the relations
(see Eqs. (\ref{eq:326c}) and (\ref{eq:326d}))
\begin{eqnarray}
x\frac{d}{dx}K^0_b =\frac{\pi\alpha^2}{m\lambda}K^0_c,
\hspace{15pt}x\frac{d}{dx}K^0_c=0,
\end{eqnarray}
we can readily deduce from the integral equations
(\ref{eq:327})-(\ref{eq:329}) the following relationships
between these quantities:
\addtocounter{equation}{1}
\setcounter{var}{\value{equation}}
\renewcommand{\theequation}
{\arabic{section}.\arabic{var}\alph{equation}}
\setcounter{equation}{0}
\begin{eqnarray}
x\frac{d\bar{\Gamma}^0}{dx}&=&\frac{\pi\alpha^2}{m}
\tilde{\Gamma}_{cross},
\label{eq:333a}\\
x\frac{d\tilde{\Gamma}_{cross}}{dx}&=&\frac{2\pi\alpha^2}
{m}\tilde{\Gamma}_{bubble}\tilde{\Gamma}_{cross},
\label{eq:333b}\\
x\frac{d\tilde{\Gamma}_{bubble}}{dx} &=& -\frac{m}{4\pi}+
\frac{\pi\alpha^2}{m}(\tilde{\Gamma}_{bubble})^2.
\label{eq:333c}
\end{eqnarray}
\setcounter{equation}{\value{var}}
\renewcommand{\theequation}
{\arabic{section}.\arabic{equation}}
\hspace {-13pt}
Note that the first term in the right hand side of
(\ref{eq:333c}) originates from the cutoff dependence
of the leading-order amplitude $ \int K^0_c (i\Delta)
(i\Delta)K^0_c $ from $ \Gamma_{bubble} $ (see
(\ref{eq:329})).
Solving the differential equation (\ref{eq:333c}),
we have 
\begin{eqnarray}
\tilde{\Gamma}_{bubble}=\left(\frac{m}{2\pi|\alpha|}
\right)\frac{1-(d_1 x)^{|\alpha|}}{1+ (d_1 x)^{|\alpha|}},
\label{eq:334}
\end{eqnarray}
where $ d_1 $ is a certain expression which is
independent of $ x. $
Using Eq. (\ref{eq:334}) with Eq. (\ref{eq:333b}),
we also find
\begin{equation}
\tilde{\Gamma}_{cross}=d_2\frac{(d_1 x)^{|\alpha|}}{\{1+
(d_1 x)^{|\alpha|}\}^2}
\label{eq:335}
\end{equation}
and then, from Eq. (\ref{eq:335}) and Eq. (\ref{eq:333a}),
\begin{eqnarray}
\bar{\Gamma}^0=-\left(\frac{\pi|\alpha| d_2}{m}\right)
\frac{1}{1+(d_1 x)^{|\alpha|}}+d_3, \label{eq:336}
\end{eqnarray}
where we introduced two $ x $-independent integration
constants $ d_2 $ and $ d_3 . $ Actually, by a simple
dimensional reason, $ d_2 $ and $ d_3 $ may depend on
$ \alpha $ only while $ d_1 $ can be put as
\begin{equation}
d_1=\left[\frac{e^{i\pi/2}}{p}f(\alpha)\right]^2
\label{eq:337}
\end{equation}
($ p $ is the magnitude of the {\em relative}
momentum), with $ f(\alpha)=1+\cal{O}(\alpha) $
on the basis of the result in lowest non-trivial
order.

Inserting the expressions (\ref{eq:334}) and
(\ref{eq:335}) into Eq. (\ref{eq:331}) yields the 
following expression for $ \Gamma_{contact} $:
\begin{eqnarray}
\Gamma_{contact} &=& \lambda d_2\frac{\frac{(d_1 
x)^{|\alpha|}}{1+(d_1 x)^{|\alpha|}}}{1-\frac{\lambda m}
{2\pi|\alpha|}+(1+\frac{\lambda m}{2\pi|\alpha|})(d_1 x)^{
|\alpha|}} \label{eq:338}
\end{eqnarray}
The desired full s-wave amplitude, given by
Eq. (\ref{eq:330}), follows immediately from
Eqs. (\ref{eq:336}) and (\ref{eq:338}).
Since the s-wave two-particle scattering amplitude can be
identified with $ -2i\Gamma^0, $ we now see that our
field-theoretic analysis leads to 
\begin{eqnarray}
A_s (p)&=&-2id_3 (\alpha)-\frac{2\pi i|\alpha|}{m}d_2 
(\alpha)\frac{\lambda-\frac{2\pi|\alpha|}{m}}
{\frac{2\pi|\alpha|}{m}-\lambda+(\frac{2\pi|\alpha|}{m}
+\lambda)(d_1 x)^{|\alpha|}} \nonumber \\
&=&-2id_3 (\alpha)-\frac{2\pi i|\alpha|}{m}d_2 (\alpha)
\frac{\lambda-\lambda_0}{\lambda_0 -
\lambda+(\lambda_0 +\lambda)[\frac{\Lambda}{p}e^{i\pi/2}
f(\alpha)]^{2|\alpha|}}, \label{eq:339}
\end{eqnarray}
where, in the second expression, $ \lambda_0 =\frac{
2\pi |\alpha|}{m} $ and we have made use of the form
(\ref{eq:337}). Based on this expression, we notice
that if $ \lambda=\pm\lambda_0=\pm\frac{2\pi |\alpha|}
{m}(=\pm\frac{e^2}{m\kappa}), $ any dependence on the
ultraviolet cutoff $ \Lambda $ disappears from $ A_s(p) $
and the resulting amplitudes exhibit {\em scale
invariance}. (just as was the case with the quantum
mechanical expression for those $ \lambda $-values)
For these special values of $ \lambda, $ the scattering
amplitude is given by a finite perturbation
theory and the system requires no renormalization.
This happens because divergences
appearing in the perturbation theory of $ \bar{\Gamma}^0 $
get cancelled order by order by divergences resulting
from contributions involving the contact interaction
\cite{sk}.
Now, at least for these finite-theory cases, we may
be allowed to invoke the usual one-to-one correspondence
existing between a nonrelativistic quantum field theory
and the quantum mechanical approach \cite{fw} ---viz.,
for $ \lambda=\pm\lambda_0 , $ our amplitude (\ref{eq:339})
should match the result in Eq. (\ref{eq:230}). [We have not
been able to verify this assertion directly, however.]
Taking this for granted, we can now fix $ d_3 (\alpha) $
and $ d_2 (\alpha) $ in the above expression by using
the result in (\ref{eq:230}) as
\addtocounter{equation}{1}
\setcounter{var}{\value{equation}}
\renewcommand{\theequation}
{\arabic{section}.\arabic{var}\alph{equation}}
\setcounter{equation}{0}
\begin{eqnarray}
-2id_3 (\alpha)&=&e^{-i\pi |\alpha|}-1,\\
-2id_3 (\alpha)&+&\frac{2\pi i |\alpha|}{m}
d_2 (\alpha)=e^{i\pi |\alpha|}-1,
\end{eqnarray}
\setcounter{equation}{\value{var}}
\renewcommand{\theequation}{\arabic{section}.\arabic{equation}}
\hspace {-13pt}
and this leads to the following expression for $ A_s (p)
$:
\begin{eqnarray}
A_s (p)=e^{-i\pi|\alpha|}-1-(e^{i\pi|\alpha|}-
e^{-i\pi|\alpha|})\frac{\lambda-\lambda_0}{\lambda_0 -
\lambda+(\lambda_0 +\lambda)[\frac{\Lambda}{p}e^{i\pi/2}
f(\alpha)]^{2|\alpha|}} \label{eq:341}
\end{eqnarray}

For $ \lambda\neq\pm\lambda_0, $ we must renormalize the
theory to obtain the scattering amplitude which has no
explicit dependence on the cutoff. That is, we will
regard $ \lambda $ as a bare coupling and elect to
introduce $ \lambda_{ren}, $ the renormalized coupling,
by the relation
\begin{eqnarray}
\frac{\lambda_0 +\lambda}{\lambda_0-\lambda}(\Lambda 
f(\alpha))^{2|\alpha|}=\frac{\lambda_0 +\lambda_{ren}}
{\lambda_0-\lambda_{ren}}\mu^{2|\alpha|}.
\label{eq:342}
\end{eqnarray}
Here, $ \mu $ is the normalization scale. Then the
amplitude (\ref{eq:341}) can be recast into the form
\begin{eqnarray}
A_s (p)= e^{-i\pi|\alpha|}-1 -(e^{i\pi|\alpha|}-
e^{-i\pi|\alpha|})\frac{\lambda_{ren}-\lambda_0}
{\lambda_0-\lambda_{ren}+(\lambda_0 +\lambda_{ren})
(\mu/p)^{2|\alpha|}e^{i\pi|\alpha|}} \label{eq:343}
\end{eqnarray}
This is in complete agreement with the quantum mechanical
expression (\ref{eq:230}), only if our field theoretic
renormalized coupling $ \lambda_{ren} $ is taken to be
related to the self-adjoint extension parameter $ \theta $
by Eq. (\ref{eq:229}). We may now assert that the quantum
field theory defined through the action (\ref{eq:31})
and renormalized as above provides an equivalent
description of many anyon quantum mechanics with a general
boundary condition as considered in the previous section.

Before closing this section, we shall emphasize once again
the role of the contact interaction terms in the quantum
description of anyons. As was explained in Sec. \ref{se:2},
we need them to implement dynamically (i.e., through the
Hamiltonian) a suitable boundary condition at the
two-anyon coincidence point. As such, their presence is
in no way an artifact of perturbation theory.
They essentially go over to the field theoretic
description, where one usually does not consider the
boundary condition separately. In fact, without including
the contact term in the Lagrangian density, the given
field theory is not renormalizable and hence does not
lead to a well-defined theory. The equivalence between
the first- and second-quantized approaches can be
established only when we include the appropriate contact
interaction term. For instance, in the special case of
anyons satisfying the hard-core boundary condition, the
Lagrangian density of the corresponding field theory
reads
\begin{eqnarray}
{\cal L}= \frac{\kappa}{2}{\partial}_t {\bf A} \times
{\bf A}- \kappa A_0 B+\phi^\dagger (i D_t +
\frac{{\bf D}^2}{2m})
\phi - \frac{e^2}{2m\kappa} \phi ^\dagger \phi ^\dagger
\phi \phi ,
\end{eqnarray}
and this happens to be an ultraviolet finite theory.

\section{SUMMARY AND DISCUSSIONS}
\label{se:4}

Due to the singular nature of anyon interaction at short
distance, one has a well-defined many anyon quantum
mechanics only after a suitable boundary condition at
the two-anyon coincidence point has been chosen.
This introduces a new free parameter---the self-adjoint
extension parameter $ \theta $ in the theory, which
serves to specify the chosen boundary condition.
We have shown that this boundary condition, in its full
generality, can be implemented dynamically by introducing
a suitable contact interaction term in the anyon
Hamiltonian. This system admits a quantum field theoretic
description in the form of a Chern-Simons gauge theory,
and we have here shown that the $ \phi^\dagger
\phi^\dagger \phi \phi $-type contact interaction assumes
a crucial role not only in securing a well-defined theory
but also in realizing the full equivalence with the quantum
mechanical approach. The strength of the renormalized
contact coupling in the field-theoretic description is
related to the self-adjoint extension parameter which
encodes the quantum mechanical boundary condition.

We have a few comments to make. First of all, note that
more general kinds of anyons (other than the ones we
discussed here) are possible, such as those obeying
so-called matrix(or mutual) statistics \cite{wz} and 
also those obeying non-Abelian statistics \cite{ww}.
Both quantum-mechanical and field-theoretic descriptions
for these generalized (but still nonrelativistic) anyons
were discussed by various authors \cite{klk,kl} without 
paying due attention to the contact interaction terms. 
These must be corrected along the line discussed in this
paper. [In this regards, see especially Ref. \cite{bb}
where the related issue is studied in non-Abelian
Chern-Simons field theory with the help of some 
lower-order perturbative calculations.]
Another problem deserving more study is to look at
related issues from the viewpoint of {\em relativistic}
Chern-Simons field theory, as was considered recently in
Ref. \cite{bfp} whithin one-loop approximation.
Finally, we need to have more information on those
specific features of an anyon system which depend crucially
on the self-adjoint extension parameter(or, equivalently,
on the coupling strength of the $ \phi^\dagger
\phi^\dagger \phi \phi $ interaction in the field theoretic
approach). After all, if anyon play a role in real physical
phenomena, it will be an experimental question to determine
what specific boundary condition the given anyons satisfy.

\begin{acknowledgments}
We would like to thank Dr. C. Kim for collaboration
on this work at the initial stage. This work was
supported in part by the Korea Science and Engineering
Foundation (through the Center for Theoretical Physics,
SNU) and the Basic Science Research Institute Program
(Project No. BSRI-96-2418), Ministry of Education,
Korea.
\end{acknowledgments}
\newpage
\centerline{\large{\bf FIGURECAPTIONS}}
\begin{itemize}
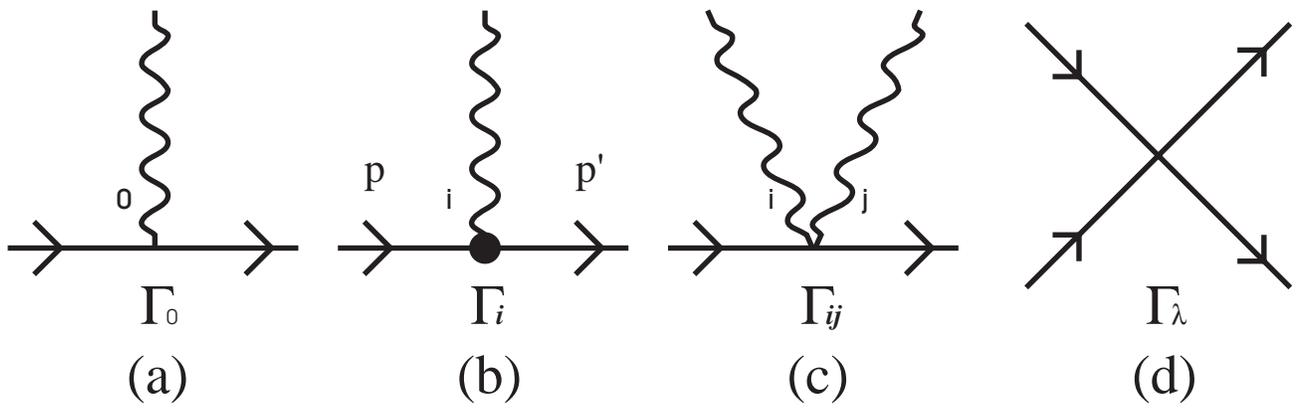
\item[FIG. 1.] Basic vertices of the theory defined by the
action (3.1)
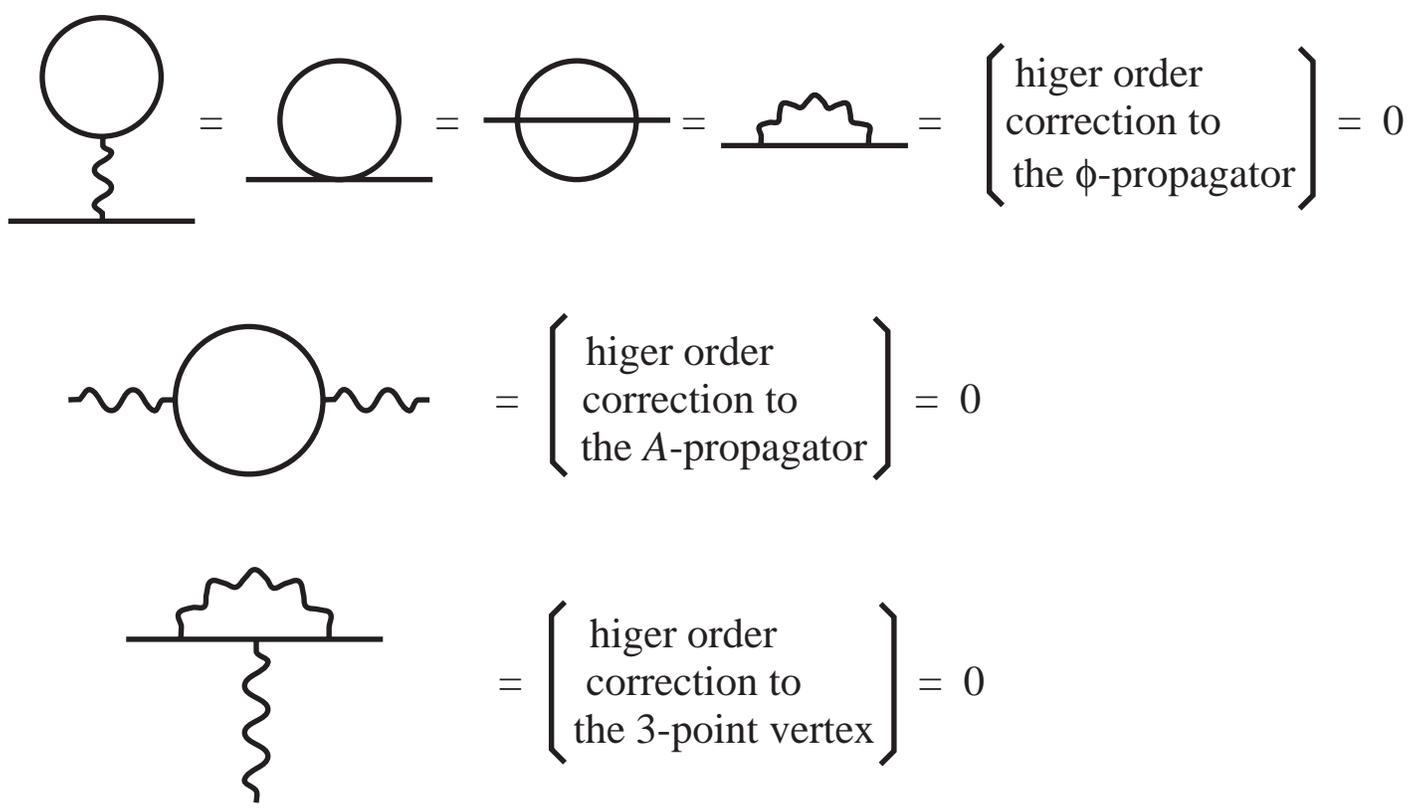
\item[FIG. 2.] Vanishing diagrams
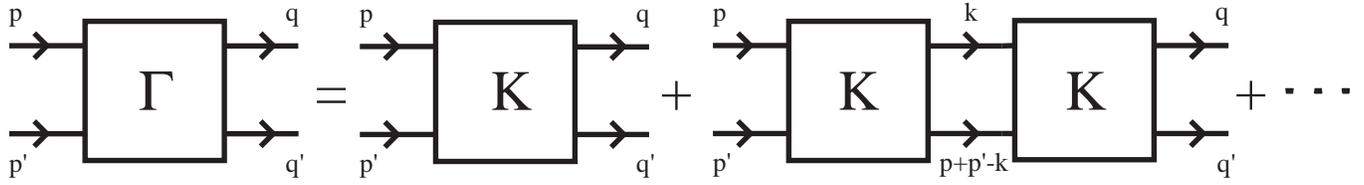
\item[FIG. 3.] Graphical representation of the effective
two-particle interaction $ \Gamma $
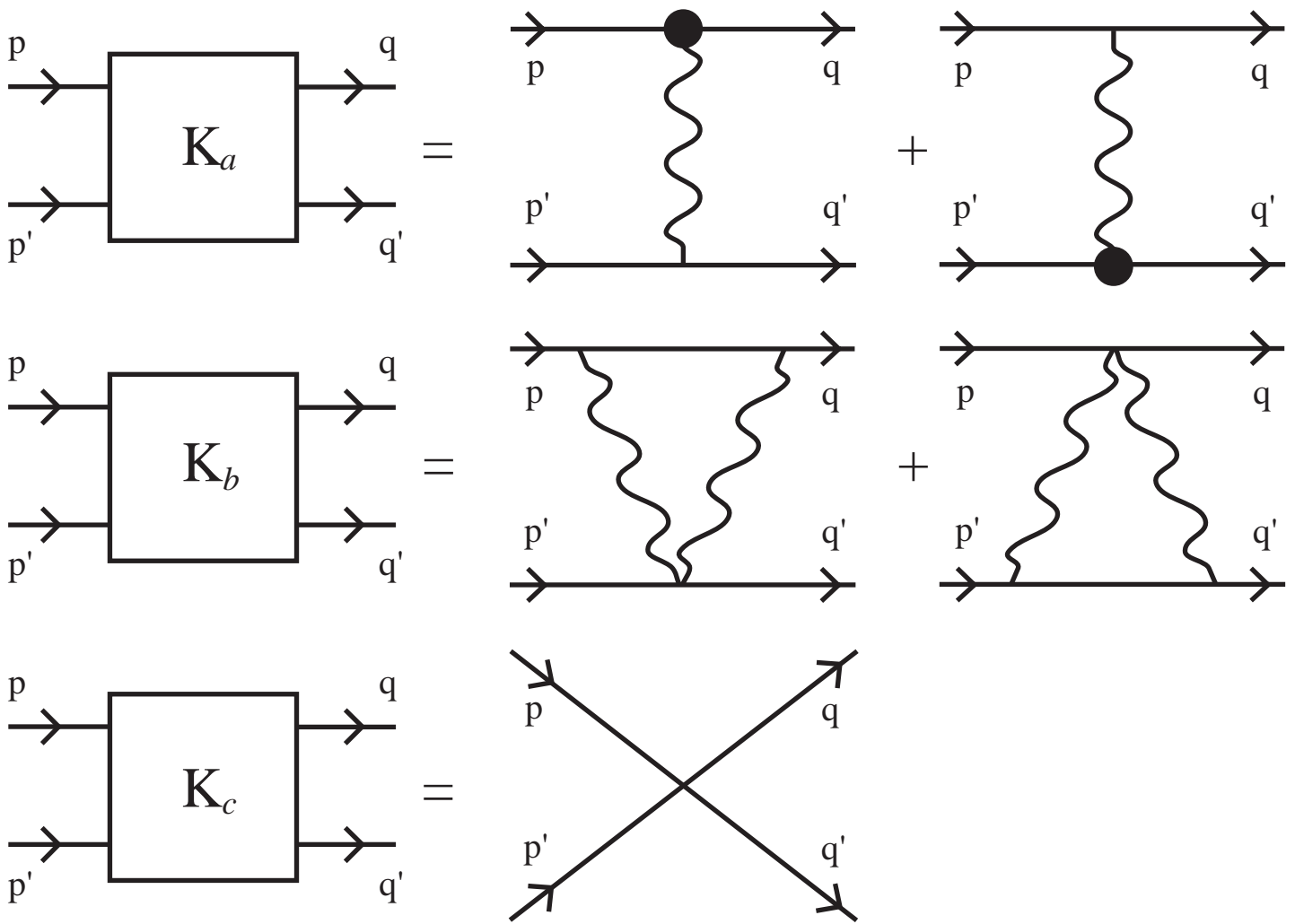
\item[FIG. 4.] Representation of the full kernel K in terms
of three different contributions
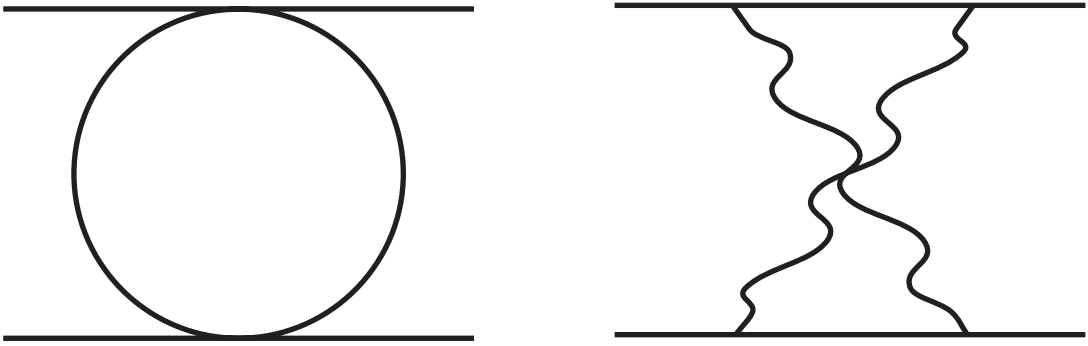
\item[FIG. 5.] Diagrams yielding vanishing contribution to
the two-particle irreducible kernel
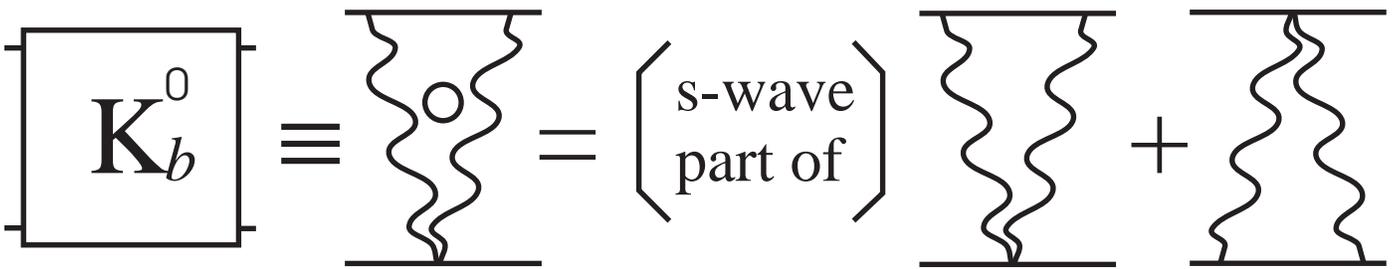
\item[FIG. 6.] Graphical representation of $ K^0 _b $
\item[FIG. 7.]Graphical representation of $ \bar{\Gamma}^0 $
, $ \Gamma_{cross} $ and $ \Gamma_{bubble} $
\end{itemize}

\end{document}